\documentclass[aps, pre, twocolumn, superscriptaddress, reprint]{revtex4-2}
\usepackage[T1]{fontenc}
\usepackage{times}
\usepackage{xfrac}
\usepackage{xcolor}
\usepackage{graphicx}
\usepackage{bm}
\usepackage{amsfonts, amsmath,amssymb, bm, graphicx}
\usepackage{siunitx}
\usepackage[english]{babel}
\usepackage{lipsum}
\usepackage{upgreek}
\usepackage{soul}
\usepackage{physics}
\usepackage{changes}
\usepackage[colorlinks=true,allcolors=blue]{hyperref}
\usepackage{footmisc}

\newcommand{\figref}[2]{\hyperref[#1]{\ref{#1}(#2)}}

\begin{document}

\makeatletter
\def\frontmatter@thefootnote{%
 \altaffilletter@sw{\@fnsymbol}{\@fnsymbol}{\csname c@\@mpfn\endcsname}%
}%

\makeatother
\title{Effective Theory of Ultrafast Skyrmion Nucleation}

\author{Rein Liefferink}\email{rein.liefferink@ru.nl}
\affiliation{Radboud University, Institute of Molecules and Materials, Heyendaalseweg 135, 6525 AJ Nijmegen, The Netherlands}

\author{Lukas K\"orber}
\affiliation{Radboud University, Institute of Molecules and Materials, Heyendaalseweg 135, 6525 AJ Nijmegen, The Netherlands}

\author{Kathinka Gerlinger}
\affiliation{Max Born Institute for Nonlinear Optics and Short Pulse Spectroscopy, 12489 Berlin, Germany}

\author{Bastian Pfau}
\affiliation{Max Born Institute for Nonlinear Optics and Short Pulse Spectroscopy, 12489 Berlin, Germany}

\author{Felix Büttner}
\affiliation{Helmholtz-Zentrum Berlin für Materialien und Energie, 14109 Berlin, Germany}
\affiliation{Experimental Physics V, Center for Electronic Correlations and Magnetism, University of Augsburg, 86159 Augsburg, Germany}

\author{Johan H. Mentink}\email{johan.mentink@ru.nl}
\affiliation{Radboud University, Institute of Molecules and Materials, Heyendaalseweg 135, 6525 AJ Nijmegen, The Netherlands}

\date{\today}

\begin{abstract}
Laser-induced ultrafast skyrmion nucleation has been experimentally demonstrated in several materials. So far, atomistic models have been used to corroborate experimental results. However, such simulations do not provide a simple intuitive understanding of the underlying physics. Here, we propose a coarse-grained effective theory where skyrmions can be nucleated or annihilated by thermal activation over energy barriers. Evaluating these two processes during a heat pulse shows good agreement with atomistic spin dynamics simulations and experiments while drastically reducing computational complexity. Furthermore, the effective theory provides a direct guide for experimentally optimizing the number of nucleated skyrmions. Interestingly, the model also predicts a novel pathway for ultrafast annihilation of skyrmions. Our results pave the way for a deeper understanding of ultrafast strongly nonequilibrium dynamics in magnetism and can generalize to ultrafast phase transitions in other long-range ordered systems.
\end{abstract}

\maketitle

\emph{Introduction ---}
Pushing the limits of controlling matter at ever smaller length and time scales is a central theme across physics today, driving numerous groundbreaking experimental discoveries \cite{kirilyuk_ultrafast_2010,stojchevska_ultrafast_2014,stoica_optical_2019,de_la_torre_colloquium_2021,johnson_ultrafast_2022}. These developments challenge established theoretical frameworks that are often valid only near thermodynamic equilibrium. In particular, in ultrafast physics, many conventional approximations break down due to the presence of strong fluctuations, reduced correlations, and multiscale effects. 

In magnetism, one of the few approaches that reliably captures such far-from-equilibrium behavior is atomistic spin dynamics (ASD), which explicitly evolves millions of coupled spins. While these simulations have yielded essential microscopic insights, it remains challenging to extract general principles, identify controlling parameters, or develop a predictive intuition for different ultrafast phenomena. Phenomenological theories have proven instrumental in such contexts. For example, effective theory \cite{radu_transient_2011,Mentink2012} was crucial in uncovering the mechanism of single-pulse all-optical switching (AOS) in ferrimagnets \cite{stanciu2007}, extending it to many materials \cite{schellekens_microscopic_2013,davies_pathways_2020,jakobs_universal_2022} and enabling nanoscale control \cite{graves_nanoscale_2013,Gorchon2016_Role_of_electron,Atxitia2018,iacocca_spin-current-mediated_2019,ksenzov_nanoscale_2021,steinbach_accelerating_2022,van_hees_toward_2022,steinbach2024exploring,yao2022all}. However, this sheds only limited light on ultrafast dynamics in ferromagnets, where the AOS mechanism has no straightforward analogue by lack of a local multi-sublattice degree of freedom. Ultrafast control of ferromagnets therefore remains a long-standing challenge.

Skyrmions provide an alternative entry point to explore ultrafast physics of ferromagnets. These remarkably stable, swirling magnetization textures can be localized within nonequilibrium correlation lengths and persist even in the presence of strong fluctuations. Their ultrafast dynamics under strong thermal fluctuations were evidenced by recent experiments which demonstrated laser-induced skyrmion creation on the picoseconds time scale \cite{Je2018,Berruto2018,Buettner2020,Gerlinger2022,li_room-temperature_2024}. Moreover, ASD simulations indicate controlled nucleation/annihilation of single skyrmions under localized heating \cite{Koshibae2014,wang_efficient_2024}. A key property of skyrmions is their magnetization profile $\bm{m}(\bm{r})$ with unit topological charge
\begin{equation}
    N = \frac{1}{4\pi}\int\limits_S \bm{m}\cdot \left(\frac{\partial\bm{m}}{\partial x}\times\frac{\partial\bm{m}}{\partial y}\right)\ \mathrm{d}x\mathrm{d}y=\pm 1,
\end{equation}
where the sign depends on the skyrmion type and the orientation along the $z$-axis (assuming a skyrmion in the $xy$-plane). When the integration area $S$ contains multiple skyrmions, $N$ also counts their total number.

As with AOS, an effective theory of ultrafast skyrmion nucleation and decay could play a key role in explaining the effect and making predictions for new experiments. By reducing the dynamics to a few physically transparent parameters, such a theory provides intuition, enables efficient exploration of parameter space, and offers a practical bridge between the current ASD results and experiment \cite{Buettner2020}.

The most notable existing effective theories of skyrmion dynamics feature collective variables approaches within micromagnetism, the continuum counterpart of ASD. A prominent example is the Thiele equation, which emerges by assuming a rigid skyrmion profile $\bm{m}(\bm{r},t)=\bm{m}_0(\bm{r}-\bm{X}(t))$ parametrized by a center-of-mass coordinate $\bm{X}(t)=(x(t),y(t))$, and integrating out all other degrees of freedom  \cite{Thiele1973,Weissenhofer2021}. The obtained equation describes movement of skyrmions by effective forces due to repulsion from a boundary or by other skyrmions and drag by field gradients, spin-polarized currents, or scattering with magnons \cite{Garst2014}. Despite their wide applicability, these particle-like theories describe comparatively slow (nanosecond) dynamics and conserve the skyrmion number $N$, and therefore do not support skyrmion nucleation or annihilation.

\begin{figure}[t]
    \includegraphics{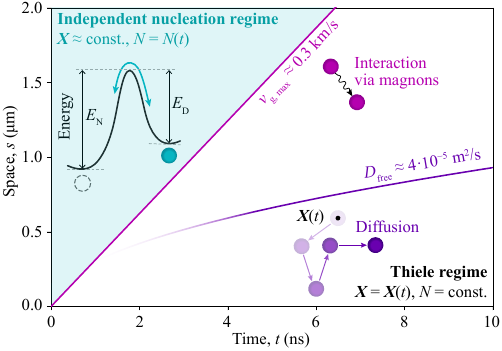}
    \caption{Space-time diagram of skyrmion dynamics illustrating the regime of validity of the presented theory for nucleation and decay of non-interacting skyrmions. Slow dynamics like changes in their position $\bm{X}(t)$ by free diffusion are governed by the Thiele equation that preserves the skyrmion number $N$. Long-range interactions between skyrmions are limited by the maximum group velocity $v_\mathrm{g,max}$ of the magnons that can efficiently scatter with skyrmions (see text for the estimation of $D_\mathrm{free}$ and $v_\mathrm{g,max}$). Thus, outside of the magnon``light cone'', the positions $\bm{X}\approx \mathrm{const}.$ and nucleation/decay dynamics of $N(t)$ can be assumed as independent activation events over an energy barrier.}
    \label{fig:regimes}
\end{figure}

In this Letter, we propose a new effective theory for ultrafast skyrmion dynamics in the regime complementary to that of the Thiele equation (colored region in Fig.~\ref{fig:regimes}).
Here, the key dynamical variable is the skyrmion number $N=N(t)$, while the skyrmion positions $\bm{X}_i$ are approximately fixed. In this ultrafast regime, the system dominantly evolves through discrete nucleation and annihilation events rather than continuous motion. Crucially, these events thus occur before significant skyrmion motion or interactions play a role, allowing us to isolate the dominant local physics. Treating the fluctuations as spatially localized then enables the use of established statistical methods for independent particles, leading to a simple description of ultrafast nonequilibrium skyrmion dynamics. The resulting theory not only captures previously reported results of skyrmion nucleation by a heat pulse, but also predicts new laser-induced annihilation scenarios, in excellent agreement with atomistic simulations.

The remainder of this letter is organized as follows. First, we provide estimates for interaction pathways between skyrmions. We use this to map out a significant space-time region where nucleation and decay events are approximately independent, defining the regime of validity of the theory. Second, we derive a kinetic theory for the skyrmion number $\expval{N(t)}$ and analyze its behavior for different temperature profiles. Third, we confirm these predictions with atomistic simulations. Finally, we discuss implications and conclusions.

In the absence of strong thermal gradients, field gradients, or spin currents, skyrmion motion is dominated by thermal diffusion. Breathing and other vibrational modes may also be thermally excited, but are neglected as they do not contribute to center-of-mass motion. As in most skyrmion-hosting materials significant pinning occurs \cite{gruber_skyrmion_2022,zazvorka_thermal_2019,Brems2025}, we first consider idealized free diffusion of a high-quality low-damping material to obtain an upper bound on skyrmion diffusion. Typical material parameters are a Gilbert-damping factor on the order of $\alpha_\mathrm{G} = 0.01$,  sample magnetic layer thickness around $d=\SI{10}{\nano\meter}$ and saturation magnetization around $M_\mathrm{s}=\SI{1.1}{\mega\ampere/\meter}$ at room temperature $T=\SI{300}{\kelvin}$ \cite{Buettner2018,sampaio_nucleation_2013}. The coefficient of free diffusion can then be calculated using the Thiele equation. In the Belavin-Polyakov limit \cite{Miltat2018,Weissenhofer2021,Belavin1975}, we can approximate it as (details in Supplemental Material), for details on the skyrmion diffusion coefficient, the estimation of the maximum magnon velocities, the fitting of the nucleation and decay times from ASD simulations, and a representative calculation of the skyrmion nucleation and decay energie barriers, which includes references \cite{Belavin1975,Garst2014,Schuette2014,Rohart_path_to_collapse_2016,Miltat2018,Bessarab2018,Buettner2018,Weissenhofer2021})
\begin{equation}
    D_\text{free}=k_\text{B} T\frac{1}{G}\frac{\alpha_\mathrm{G}}{1+\alpha_\mathrm{G}^2}\approx 4\cdot 10^{-5}\text{\,m$^2$/s}.
\end{equation}
Note that this approach breaks down at very short time scales since the effects of defects become larger and magnons and precession come into play (e.g. Walker breakdown). This calculation indeed provides an upper bound for skyrmion diffusion, as experimentally determined values are orders of magnitude lower \cite{zazvorka_thermal_2019,Brems2025}.

The corresponding space-time regime is shown in Fig.~\ref{fig:regimes}, being limited by the distance $s=\sqrt{2{D}_\text{free}t}$ covered during free diffusion. Outside of this regime, short-range interactions (e.g. repulsion) between different skyrmions are strongly suppressed. Long-range interactions mediated by magnons are, in return, limited by their maximum group velocity $v_{\mathrm{g,max}}$. As shown in the Supplemental Material, we estimate an upper bound for the velocity of magnons that can scatter with skyrmions and therefore mediate long-range interactions as
\begin{equation}
    v_{\text{g},\,\max}= 8\gamma\frac{A_\text{ex}}{M_\text{s}}\frac{1}{R}\approx 0.3\text{\,km/s},
\end{equation}
where we assert $A_\text{ex}=\SI{12}{\pico\joule/m}$ as a typical micromagnetic exchange stiffness constant and $R\approx\SI{50}{\nano\meter}$ as the skyrmion radius \cite{Buettner2018, sampaio_nucleation_2013, Buettner2020}.

As shown in Fig.~\ref{fig:regimes}, the magnon velocity thus defines a region of effectively space-like separation outside of the magnonic light cone $s=v_{\text{g},\,\max}t$ where the skyrmion positions $\bm{X}\approx \mathrm{const.}$ and interaction between skyrmions is negligible. We note that magnons in typical skyrmion materials can be strongly attenuated, effectively enlarging the non-interacting regime.

\emph{Effective theory ---}
We proceed to construct a kinetic equation for the topological charge $N(t)$ that can only change through nucleation or annihilation of individual skyrmions. Within the aforementioned space-time regime of non-interacting skyrmions, these nucleation and annihilation events can be treated as statistically independent, allowing substantial simplifications.

At finite temperature, the number of skyrmions $N$ is a stochastic quantity. Therefore, we sum over all possible microstates $i$ (i.e. configurations of skyrmions) to obtain the ensemble average
\begin{equation}
    \expval{N}  = \sum\limits_{i\in \text{all microstates}} P_i N_i
\end{equation}
with $P_i$ and $N_i$ respectively being the probability of finding the system in the microstate $i$ and the number of skyrmions in this microstate. We can obtain microstates using the discussed independence by coarse-graining the sample into $N_\mathrm{max}$ statistically equivalent regions such that each one can host exactly one skyrmion.

It then follows that a microstate $i$ with $N$ skyrmions has probability $P_i=p^N (1-p)^{N_\text{max}-N}$, where $p$ is the chance to find a skyrmion in a single region. Microstates with $N_i=N$ are obtained by choosing $N$ regions with a skyrmion out of $N_\text{max}$, which can be done in $\binom{N_\text{max}}{N}$ ways. Hence,
\begin{equation}
    \expval{N}  = \sum_{N=0}^{N_\text{max}} \binom{N_\text{max}}{N} p^N (1-p)^{N_\text{max}-N} N = pN_\text{max}.
\end{equation}
In particular, the full distribution of the stochastic variable $N$ is binomial with probability $p$ over $N_\text{max}$ samples. This may be expected since we are dealing with $N_\text{max}$ independent binary systems with probabilities $p$. The problem is now reduced to finding a kinetic equation for the local variable $p$.

We focus on the simplest reported case where skyrmions are the only observed magnetic texture \cite{Buettner2020,Je2018,li_room-temperature_2024}. Then a given skyrmion-supporting region can be either in a homogeneous magnetization state or skyrmion state. In a time interval $\text{d} t$, transitions between these states in either direction can lead to a change $\text{d} p$ of $p$. We can write
\begin{equation}
    \text{d} p=(\text{d} p)_\text{nuc}-(\text{d} p)_\text{dec},
\end{equation}
where $(\text{d} p)_\text{nuc}$ and $(\text{d} p)_\text{dec}$ are the respective probabilities of transition towards and from the skyrmion state, that is a nucleation or decay event, within $\text{d} t$. A skyrmion with lifetime $\tau_\text{D}$ has a chance $\text{d} t/\tau_\text{D}$ to decay during $\text{d} t$. This can only happen if a skyrmion is present, which has probability $p$, so $(\text{d} p)_\text{dec}=p\,\text{d} t/\tau_\text{D}$. Nucleation is only possible in the absence of a skyrmion, leading to $(\text{d} p)_\text{nuc}= (1-p)\text{d} t/\tau_\text{N}$, where $\tau_\mathrm{N}$ is the nucleation time (or lifetime of the empty state). Assuming the independence of nucleation and decay processes, we can combine this into a differential equation
\begin{equation}
    \label{eq:p_eq_of_motion}
    \frac{\text{d} p}{\text{d} t}=\frac{1-p}{\tau_\text{N}}-\frac{p}{\tau_\text{D}}.
\end{equation}
We note that a similar equation has been derived before as a Markov master equation for transitions between (anti)skyrmions and homogenous magnetic states \cite{Goerzen_atomistic_2022,Goerzen_Lifetime_2023}.

Here we focus on transitions induced by thermal fluctuations, so time scales $\tau_\text{N}$ and $\tau_\text{D}$ are a function of temperature $T$. Temperature-driven skyrmion nucleation and especially decay have been studied in great detail, experimentally \cite{Wild2017,tang_creating_2025}, within micromagnetism \cite{Buettner2018,Bernand-Mantel2022,je_targeted_2021,verga_skyrmion_2014,Schuette2014}, atomistic spin dynamics \cite{Rozsa2016,wang_spontaneous_2021,olleros-rodriguez_non-equilibrium_2022} and transition state theory \cite{Bessarab2018,bessarab_method_2015}. The most commonly used model is Arrhenius-type activation over the energy barrier between the empty and skyrmion state \cite{Wild2017,Rozsa2016,bessarab_method_2015,Bessarab2018,desplat_paths_2019,desplat_thermal_2018,hoffmann_atomistic_2020}, in which
\begin{equation}
\label{eq:arrh}
    \tau_i=\tau_{0,i}\exp\left(\frac{E_i}{k_\text{B}T}\right)\text{ with } i=\text{N},\text{D},
\end{equation}
where $\tau_{0,\text{N}}$ and $\tau_{0,\text{D}}$ are so-called attempt times and $T$ is the temperature. The nucleation energy barrier $E_\text{N}$ is the difference between the empty state and the top of the barrier, while the decay energy $E_\text{D}$ is given by the top of the barrier compared to the skyrmion state (see Fig.~\ref{fig:regimes} inset). These two barriers together with the attempt times will constitute the only system-dependent parameters entering our effective theory.

Although corrections to Eq.~\eqref{eq:arrh} and accurate determination of its parameters remain active topics of research \cite{Buettner2018,Bernand-Mantel2022,Wild2017,Rozsa2016,bessarab_method_2015,Bessarab2018,desplat_paths_2019,desplat_thermal_2018,hoffmann_atomistic_2020}, only two fundamental features are important for our purposes. First, the strong temperature dependence due to the exponential Arrhenius form. Second, despite the nonequilibrium nature of the targeted problem, these theory parameters are features of the magnetic energy landscape that can be accessed with static calculations.

To illustrate this and show how $E_\text{N}$ and $E_\text{D}$ relate to microscopic parameters, we include a representative calculation in the spirit of \cite{Buettner2018} in the Supplemental Material. In short, we find that the nucleation energy barrier is dominated by exchange, while the decay barrier is decreased by anisotropy (and possibly exchange) but increased by Dzyaloshinskii-Moriya interaction (DMI) and dipolar interactions (given, e.g., by the
saturation magnetization). This furthermore highlights that our effective theory allows simple treatment of dipolar effects which are highly challenging in atomistic simulations.

The kinetic equation of the expectation value $\langle N\rangle=N_\text{max} p$ now directly follows by multiplying both sides of Eq.~\eqref{eq:p_eq_of_motion} with $N_\text{max}$
\begin{equation}
\label{eq:main_expvalN}
    \frac{\text{d}\langle N\rangle}{\text{d} t}=\frac{N_\text{max}-\langle N\rangle}{\tau_\text{N}} - \frac{\langle N\rangle}{\tau_\text{D}},
\end{equation}
providing the central equation of the effective theory. 

While the form of Eq.~\eqref{eq:main_expvalN} is more transparent to its physical origin, it is equivalent to the mathematically simpler
\begin{equation}
    \label{eq:exprelax}
    \frac{\text{d} \expval{N}}{\text{d} t}=\frac{N_\text{max}}{\tau_\text{N}}-\left(\frac{1}{\tau_\text{N}}+\frac{1}{\tau_\text{D}}\right) \expval{N}=-\frac{\expval{N}-N_\text{eq}}{\tau_\text{R}},
\end{equation} 
where $1/\tau_\text{R}=1/\tau_\text{D}+1/\tau_\text{N}$ and $N_\text{eq}=N_\text{max}{\tau_\text{R}}/{\tau_\text{N}}$. That is, the kinetic equation for $\expval{N}$ describes the exponential relaxation towards an instantaneous equilibrium value $N_\text{eq}$ with a response time $\tau_\text{R}$. In particular, if the rates are constant, \textit{e.g.}, at constant temperature, one has 
\begin{equation}\label{eq:analyticexponential}
    \expval{N(t)}=\expval{N(0)}+[N_\text{eq}-\expval{N(0)}]\left(1-e^{-t/\tau_\text{R}}\right).
\end{equation}

Due to its simplicity, various scenarios for ultrafast control of the number of skyrmions $\expval{N}$ emerge directly from the effective theory Eq.~\eqref{eq:exprelax}. The most obvious control is based on tuning the transient temperature profile $T(t)$, as was done for the AOS in ferrimagnets. Experimentally, temperature profiles are achieved using ultrashort laser (or current pulses for longer time scales) \cite{Je2018,Buettner2020}, which can be tuned using fluence, pulse width and sample heat dissipation properties \cite{steinbach_accelerating_2022,van_hees_toward_2022,ksenzov_nanoscale_2021,davies_pathways_2020,Gorchon2016_Role_of_electron}. 

\begin{figure}[t]
    \includegraphics{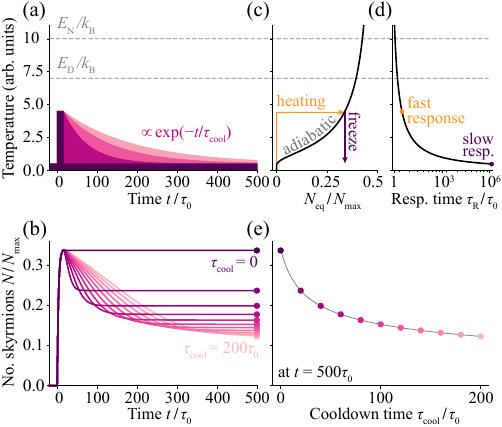}
    \caption{Calculation of heat pulse induced skyrmion nucleation dynamics $N(t)$ using the effective theory (b) for pulses with exponential cooldown times $\tau_\text{cool}$ varied between 0 and $200\tau_0$ ((a), dark purple to light pink) after an initial constant high temperature. The final values of $N$ from (b) are plotted as a function of cooldown time in panel (e). Temperature dependence of the equilibrium number of skyrmions $N_\text{eq}$ (c) and response time $\tau_\text{R}$ (c). The path of the rectangular pulse is illustrated with yellow and purple annotations. At high $T$, the response time is very short and the equilibrium number of skyrmions is enhanced, leading to nucleation. The slow response time at low $T$ locks the system into a skyrmion state after fast cooldown, while the dynamics tend towards the adiabatic path $N(t)=N_\text{eq}(T(t))$ with low final $N$ for slower cooldown.}
    \label{fig:cooldown} 
\end{figure}

As an illustration, we assert Arrhenius time scales with temperature-independent energy barriers $E_\mathrm{D}=0.7E_\mathrm{N}$ and attempt times $\tau_{\text{D},0}=\tau_{\text{N},0}=\tau_0$ for decay and nucleation, respectively. This case where the skyrmions do \textit{not} constitute the ground state, is particularly interesting, since we will see that it is nevertheless possible to generate and stabilize them. To keep the parameters of the temperature profile $T(t)$ minimal, we consider a heat pulse consisting of a rectangular stage with $T_\text{pulse}(t)=T_\text{p}=0.4\,E_\mathrm{N}/k_\mathrm{B}$ for $0<t<w_\text{p}=15\tau_0$, followed by an exponential cooling stage for $t>w_\text{p}$, given by $T_\text{pulse}(t)=T_\text{p}\exp[-(t-w_\text{p})/\tau_\text{cool}]$ and characterized by a cooling time $\tau_\mathrm{cool}$. Including a base temperature $T_0=0.05\,E_\mathrm{N}/k_\mathrm{B}$ leads to the full temperature profile $T(t)=T_0+T_\text{pulse}(t)$. Several such profiles with different values of $\tau_\mathrm{cool}$ are shown in Fig.~\ref{fig:cooldown}(a). The corresponding induced evolutions of the skyrmion number $\expval{N}$ are obtained by numerically integrating Eq.~\eqref{eq:main_expvalN} for these profiles and shown in Fig.~\ref{fig:cooldown}(b).

During the initial heating period at high constant temperature, skyrmions are rapidly nucleated and follow the simple exponential growth Eq.~\eqref{eq:analyticexponential}. Remarkably, when the temperature pulse is turned off instantly after the heating period (where $\tau_\mathrm{cool}\ll \tau_0$), the skyrmion number is "frozen in" and remains almost constant within the considered time scale. This can be understood by the strong temperature dependence of both the instantaneous equilibrium level $N_\text{eq}$ and the response time $\tau_\text{R}$ shown in Fig.~\ref{fig:cooldown}(c) and Fig.~\ref{fig:cooldown}(d), respectively. Naturally, at high temperatures, spontaneous activation over the skyrmion nucleation energy barrier becomes more likely, increasing the expected occupation of the skyrmion states $N_\textrm{eq}$. Along with the nucleation and decay time, the response time $\tau_\mathrm{R}$ is even more strongly affected by the increase in temperature, as seen in Fig.~\ref{fig:cooldown}(d). Here, the response time of the system is decreased by almost 6 orders of magnitude during the heating period, enabling $\expval{N}$ to quickly reach the new equilibrium value of $N_\text{eq}$. This explains the ultrashort timescales observed in skyrmion nucleation dynamics \cite{Buettner2020}. After turning the heat pulse off instantly, the response time conversely increases by the same orders of magnitude, effectively stopping further dynamics and freezing in the generated skyrmions.

The stabilization of non-equilibrium skyrmions after the heat pulse is only possible due to the immediate drop in temperature ($\tau_\mathrm{cool}\ll \tau_0$). If, instead, the cooling was quasi-adiabatic ($\tau_\mathrm{cool} \gg \tau_0$), the skyrmion population would always have enough time to reach the instantaneous equilibrium $N_\mathrm{eq}$ [see Fig.~\ref{fig:cooldown}(c)] and almost all generated skyrmions would be gone when the base temperature $T_0$ is reached. Indeed, gradually increasing the cooling time $\tau_\mathrm{cool}$ continuously decreases the final skyrmion population evaluated at $t=500\tau_0$, as shown in Fig.~\figref{fig:cooldown}{b} and Fig.~\figref{fig:cooldown}{e}. When the cooldown time becomes comparable to the response time at intermediate temperature $T$, a large portion of the skyrmions can annihilate before being frozen in.

The effective theory predicts that the number of nucleation skyrmions can be directly controlled using the cooldown time of the pulse, which experimentally is feasible by engineering the heat diffusion and pulse length of the laser. For example, the metastable skyrmions considered here could be written with high and short pulses, while they could be erased with lower and wider pulse profiles as a sort of annealing, representing fully heating-based stochastic switching. 

\begin{figure}[t]
    \includegraphics{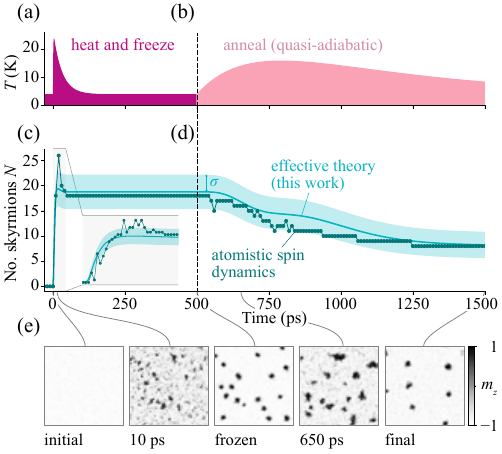}
    \caption{Prediction of (a,c) heat-induced nucleation and freezing in of skyrmions by a short heat pulse and (b,d) annihilation of the same skyrmions by slow annealing and quasi-adiabatic cooldown. (a) and (b) show the temperature profile of the different heat pulses, while (c) and (d) show the resulting mean number of skyrmions $\expval{N(t)}$ (with standard deviation $\sigma$) as predicted by the effective theory from this work, compared with a full ASD simulation. Representative snapshots of the out-of-plane magnetization $m_z$ from the ASD simulation are shown in panel (e).}
    \label{fig:ASDcomparison}
\end{figure}

To test this hypothesis and the validity of the effective theory for time-dependent temperatures, ASD simulations were performed (using the UppASD package \cite{Skubic2008}). Such simulations were previously used to corroborate experimental results on laser-induced skyrmion nucleation \cite{Buettner2020}. In this case, effective theory parameters are obtained by fitting relaxation at constant temperatures, remaining well within the range of energies expected based on the spin textures involved \cite{Schuette2014,Bessarab2018}. The details of this can be found in the Supplemental Material. In ASD, one considers discrete classical magnetic moments $\bm{m}_i$ on a lattice, usually the square lattice. The Hamiltonian for the moments is given by
\begin{align}
\label{eq:ASDham}
    \mathcal{H}_\text{ASD}=
       &-\sum_{i\neq j}J_{ij}\bm{m}_i\cdot \bm{m}_{j}
       -\sum_{i\neq j}\bm{D}_{ij}\cdot(\bm{m}_i\times \bm{m}_j)\nonumber\\ 
       &-\sum_i Km_{i,z}^2
       -\sum_iM\bm{B}\cdot \bm{m}_i,
\end{align}
which includes ferromagnetic exchange $J_{ij}>0$, DMI $\bm{D}_{ij}$, uniaxial anisotopy $K$ and the interaction with an external field $\bm{B}$. We take over the parameter set from \cite{Buettner2020} that is known to qualitatively match the experimental system, and in particular to support skyrmions. However, we increase the external field strength to $B=0.75$\,T to ensure the homogeneous ``empty'' state has a lower energy than the skyrmion state as above. We use a (spatially homogeneous) exponential rise and decay on top of a base temperature $T_0=4$\,K to model a temperature profile induced by a laser (or current) pulse: 
\begin{equation}
\label{eq:heatpulse}
     T(t) = T_0 + T_1(1- e^{-t/\tau_\text{rise}})e^{-t/\tau_\text{fall}},
\end{equation}
with $T_1=c(\tau_\text{rise},\tau_\text{fall})T_\text{peak}$, where $c$ is chosen such that the peak of the pulse is $T_\text{peak}$ above $T_0$. The parameters $\tau_\text{rise}$ and $\tau_\text{fall}$ control the rise time and decay time of the pulse, respectively. To test the temperature-induced writing and erasing, we apply two pulses as shown in Fig.~\ref{fig:ASDcomparison}(a,b) after initializing the system in the fully field-aligned state. The first pulse has height $T_\text{peak}=20$\,K with fast heating and cooling given by $\tau_\text{rise}=1$\,ps and $\tau_\text{fall}=30$\,ps, while the second pulse heats and cools more slowly with $\tau_\text{rise}=300$\,ps and $\tau_\text{fall}=500$\,ps and reduced height $T_\text{peak}=12$\,K. 

The time trace of the number of skyrmions $N$ resulting from this temperature profile is shown for an ASD simulation (points) as well as for the effective theory (solid line and shaded area) in Fig.~\ref{fig:ASDcomparison}(c,d). The solid line represents the expectation value $\expval{N}$ and the shaded area represents the standard deviation $\sqrt{\text{Var}(N)}$ of the number of skyrmions calculated using the effective theory. We find excellent agreement between the curves for $N$ for both the writing and the erasing. The dynamics of the magnetization texture in Fig.~\ref{fig:ASDcomparison}(e) are moreover in accordance with the assumptions and trends described by the effective theory. While the system is hot during the first pulse, skyrmions spontaneously appear (and disappear) at random places, most of which freeze in during the rapid cooldown. The second longer pulse drives the system closer to $N_\textrm{eq}$ by allowing a random subset of skyrmions to annihilate. 

\emph{Discussion ---}
Although we illustrated the case of metastable skyrmions, the model is equally applicable to systems in which skyrmions are the ground state ($E_\mathrm{N}\le E_\mathrm{D}$). Furthermore, within the same framework, several refinements and generalizations can be implemented with minimal changes. One refinement concerns dipolar fields (see Supplementary Materials D), which in addition to stabilization give rise to an instantaneous skyrmion-skyrmion interaction. This can be incorporated by making $E_\mathrm{D}$ density dependent, but the effect is only expected to be relevant at high skyrmion densities. Another extension is the inclusion of spatial inhomogeneity, which can be addressed by dividing the system into quasi-homogeneous regions, provided they remain larger than the skyrmion size. This is applicable to spatial intensity profiles of laser pulses, pre-existing domain patterns, and spatially inhomogeneous material parameters \cite{Kern2022}. A simple correction to the system's excitation is to include nonthermal stimuli such as external fields. Such effects are typically small compared to the ultrafast heating, but would appear as transient changes to the energy barriers $E_\mathrm{N}$ and $E_\mathrm{D}$. Finally, the framework readily accommodates more than two local magnetic states (skyrmion and empty). This involves a simple extension of the kinetic equation, in a similar vein as \cite{Goerzen_Lifetime_2023} and would enable the description of antiskyrmions and other magnetic textures.

Beyond the above generalizations, our framework supports fully interacting skyrmion systems as well, despite the approximations made. This would require solving the equations numerically, which can still be efficiently done on a discrete grid. This extension would also allow the reintroduction of skyrmion motion, bridging the ultrafast regime with the Thiele regime.

The present theory captures ultrafast skyrmion nucleation and annihilation using a minimal set of effective parameters and accurately reproduces atomistic spin-dynamics (ASD) results. The key conceptual advance is the localization approximation within the ultrafast time window, which allows individual nucleation and annihilation events to be treated as statistically independent. The success of this approximation demonstrates that the ultrafast dynamics of magnetism at the nanoscale can, to leading order, be decomposed into local processes. Our minimal formulation exposes the parameter dependencies of ultrafast skyrmion creation and decay, crucial to achieving control in experiments. Notably, we demonstrate a strong dependence on the heat dissipation rate that arises because the skyrmions are frozen in from nonequilibrium states.

Finally, the general approach of constructing coarse-grained statistical models within ultrafast time windows, where inter-grain interactions are strongly restricted, offers a way to reduce multiscale complexity not only in ultrafast magnetism \cite{Lambert2014,khusyainov_laser-induced_2024,vonKorffSchmising2014,guang2020creating,zayko2021ultrafast,khela2023laser,battistelli2024coherent} but also in ultrafast photo-induced phase transitions of other long-range-ordered systems \cite{gerasimenko_quantum_2019,johnson_all-optical_2024,padma_symmetry-protected_2025}.

\emph{Data availability ---}
Source data for figures and ASD simulation code are publicly available \cite{rein_liefferink_2026_effective_theory_dataset}. All other data that support the findings of this paper are available from the corresponding authors upon request.

\emph{Acknowledgements ---}
This work received funding from the European Union Horizon 2020 and innovation program under the European Research Council ERC Grant Agreement No. 856538 (3D-MAGiC), the EU via Horizon Europe project no. 101070290 (NIMFEIA) and by the Dutch Research Council (NWO) via VIDI project no. 223.157 (CHASEMAG). L.K. acknowledges funding by the Radboud Excellence Initiative. B.P. acknowledges funding from the Röntgen-Ångström-Cluster (Verbundprojekt 05K2024 - 2023-06359 ``Soft-XPCS''). B.P. and K.G. acknowledge funding from the Leibniz Association -- grant no. K162/2018 (OptiSPIN).
F.B. and B.P. acknowledge funding from the Deutsche Forschungsgemeinschaft (DFG) -- project no. 505818345 (Topo3D).
F.B. acknowledges funding from the Helmholtz Young Investigator Group Program (VH-NG-1520) and from DFG -- project no. 49254781 (TRR360-C2). 

\bibliography{nucdec.bib}

\end{document}